# Effects of the Electronic Structure, Phase Transition, and Localized Dynamics of Atoms in the Formation of Tiny Particles of Gold

**Mubarak Ali, [a,] * and I-Nan Lin [b]**

[a] Department of Physics, COMSATS University Islamabad, Islamabad Campus, Park Road 45550, Pakistan *E-mail: mubarak74@mail.com, mubarak60@hotmail.com

[b] Department of Physics, Tamkang University, Tamsui Dist., New Taipei City 25137, Taiwan (R.O.C.). E-mail: inanlin@mail.tku.edu.tw

**Abstract:** In addition to the self-governing properties, tiny metallic colloids are the building blocks of larger particles. This topic has been the subject of many studies. This work discusses the results of three different experiments. Attained dynamics of the atoms also play a role in developing tiny particles. Atoms at the solution surface can also bind by the nano energy packets. Arrays of atoms convert into structures of smooth elements. When electron streams impinge on gold atoms at a fixed angle, atoms can elongate further. Traveling photons along the interface affect the atomic arrays. Gold atoms can also develop different tiny particles in solution. Various factors to their development contribute. The present work also considers the analyses of some tiny-sized particles. In the processing of tiny-metallic colloids at different parameters, major leveled modifications of atoms took place. The study also discusses the influence of traveling photons along the matter-solution interface. This study highlights the fundamental process of developing a variety of tiny particles.

Several possibilities may open through the pulse-based process to develop engineered materials.

## 1.0. Introduction

Studying the matter at the nanoscale is valuable for a new beginning of the scientific era. Tiny-sized particles have the potential to revolutionize technology in different fields. Tiny metallic colloids not only develop nanoscale devices but also develop the understanding of developing higher-order materials. Tiny-metallic colloids can be a suitable candidate for several applications.

The research on tiny-metallic colloids has a long history. Only some of the studies are cited here [1-11]. Nano-sized gold particles and clusters behave like simple chemical compounds, which may find a wide range of applications in catalysis, sensors, and molecular electronics [1]. Nano-sized metallic particles collectively oscillate while trapping energetic electrons [3].

Understanding the dynamics of the development of tiny-sized particles is vital before assembling tiny-metallic colloids for a larger particle [9]. Precise control over the surface properties of nanoparticles will direct their assembly into high-order structures [11].

Tiny colloids in specific sizes can keep non-fcc geometric structures [12]. Characteristics of nanoparticles can change depending on their size [13]. Many studies in the literature discussed developing tiny particles by involving plasma-solution-based processes [14-17].

From another perspective, the ability to structure matter in the region of sub-optical wavelength can deliver unusual optical properties [18, 19]. The catalytic activity of metallic nanostructures is enhanced for controlling the phase transition [20, 21]. Visualizing and observing an atom in high resolution provides an advantage in understanding the functionalities of nanomaterials [22, 23].

Tiny particles of triangular shapes can develop at suitable pulse ON/OFF times, as discussed in the previous studies of the same authors [24-28]. Electronic structure, phase transition, and dynamics of atoms affect the developing gold tiny particles.

This study discusses the developmental process of various tiny-sized particles. The high-resolution images of those tiny particles chosen have shapes other than triangular shapes.

## 2.0. Experimental details

A pulse-based electron photon-solution interface process is a new method. In this study, the synthesis of tiny metallic colloids is by this method. By using gold (III) chloride trihydrate, a 100 ml solution was made ready in each experiment.

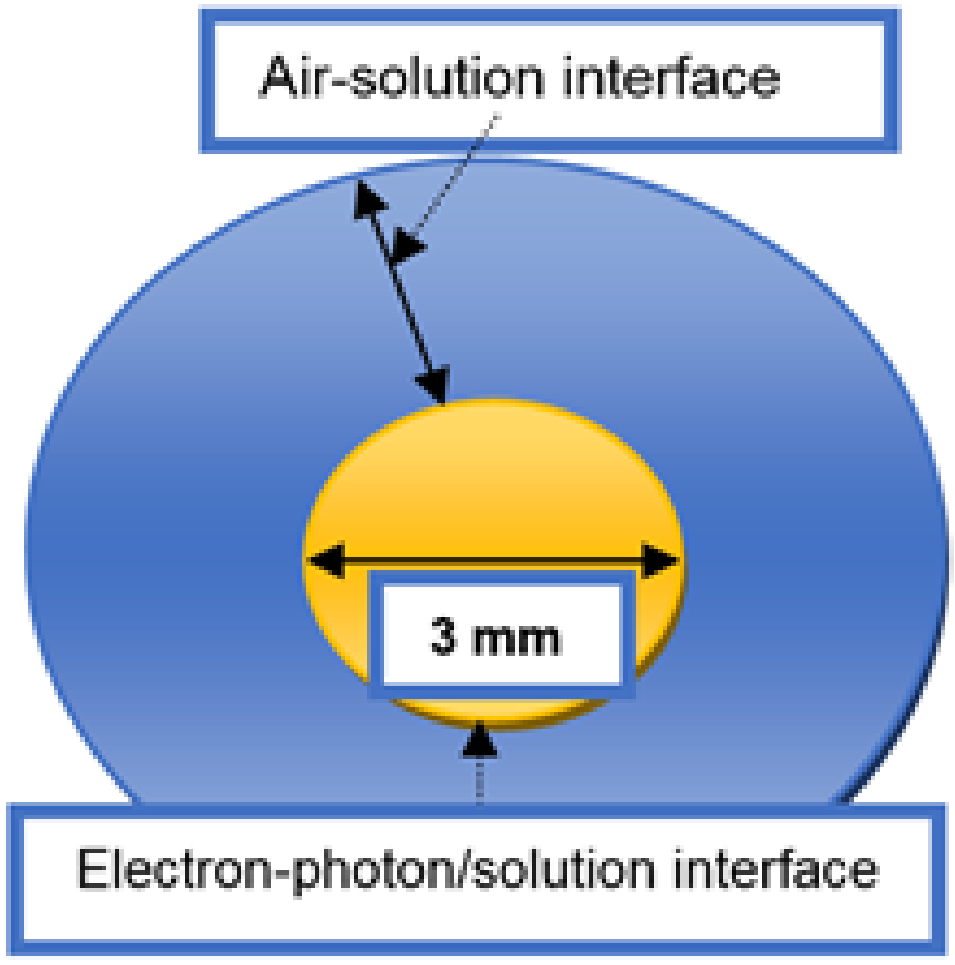

**Figure 1:** Zones of air/solution interface and electron-photon/solution interface

Figure 1 shows the zones of air/solution and electron-photon/solution interfaces. Further detail of the process is given elsewhere [24]. The high-resolution transmission optical microscopy (HR-TOM) is employed to capture the images of tiny-sized particles. It is also known as high-resolution transmission electron microscopy (HR-TEM), Model: JEOL JEM2100F.

## 3.0. Results and discussion

Figure 2 (a–c) shows images of different tiny particles taken by HR-TOM. In these tiny-sized particles, atoms got amalgamated by keeping the regular order in some portions. In Figure 2 (a-c), atoms amalgamated in non-regular order in some portions of the tiny-sized particles, too. Some atoms also deformed their shapes.

Figure 2 (d) shows different stress levels of the atoms. The covered portion in Figure 2 (d) does not show the amalgamation of atoms side by side. Therefore, those atoms do not elongate unidirectionally. As a result, atoms of a rectangular box shown in Figure 2 (d) do not convert into structures of smooth elements. In that rectangular box, where the amalgamation of the atoms was not in order, they do not show an elongation. Several atomic arrays convert into structures of smooth elements. It means the exertion of surface force at the electron levels remains even. Here, a packet of the nano energy binds the atoms of the tiny particle. In Figure 2 (d), atoms bind before their elongation.

In Figure 2 (d), atoms of the triangular-shaped tiny particle nearly keep the same elongation rate as in Figure 2 (e). Therefore, the configuration of electrons in each atom also remains the same.

In Figure 2 (e), a spherical-shaped tiny particle developed. Atoms of a tiny-sized particle have nearly the same phase as the atoms of a tiny-sized particle in Figure 2 (d). In Figure 2 (e), the width of a structure of a smooth element is ~ 0.12 nm.

In Figure 2 (e), structures of smooth elements in a tiny-sized particle also keep the same interspacing distance. In Figure 2 (e), width and interspacing distances indicate that the elongation rate of atoms remains the same. Atoms of arrays now relate to perfect structures of smooth elements. In Figure 2 (e), they look like the shape of straightened threads.

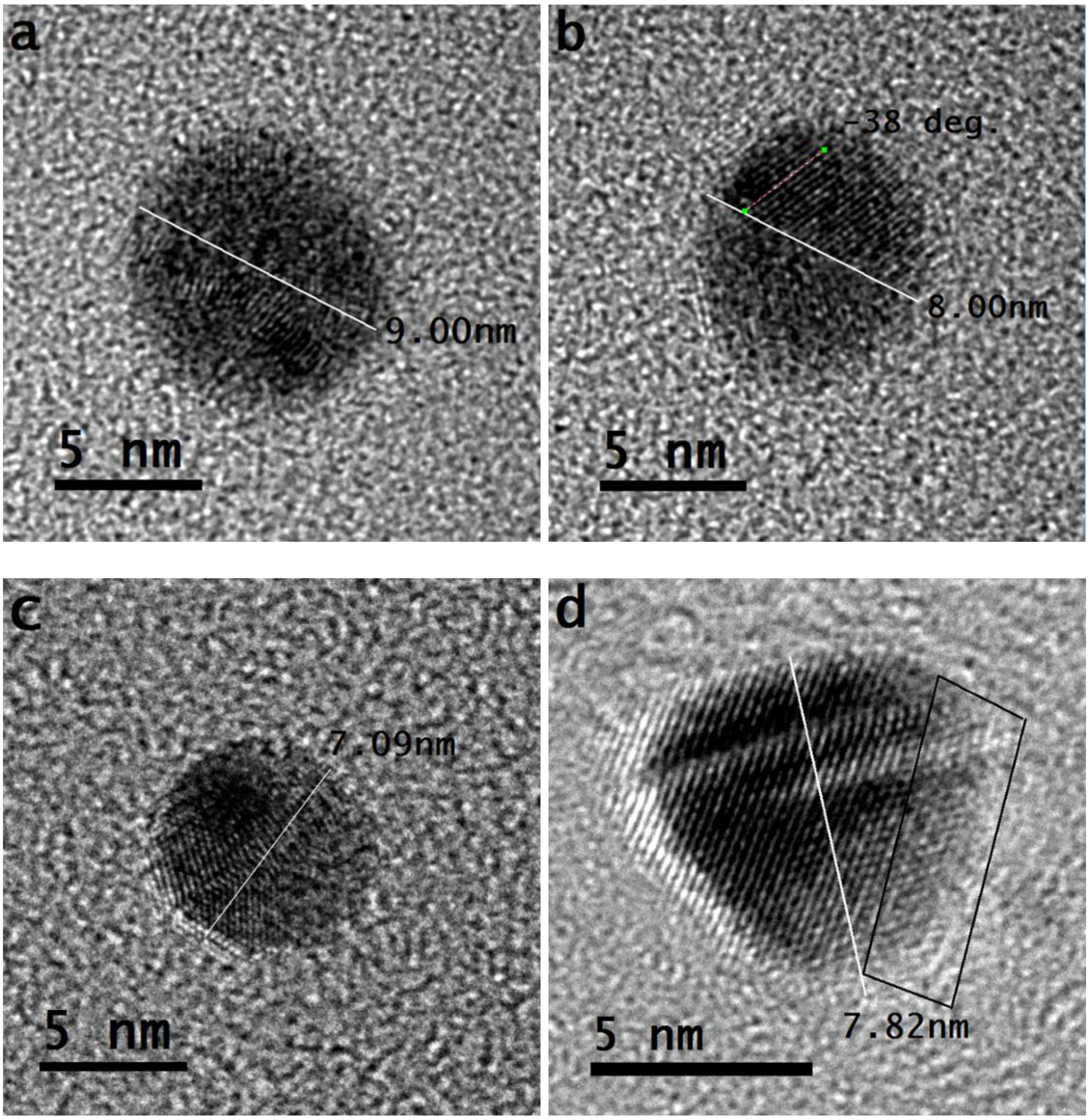

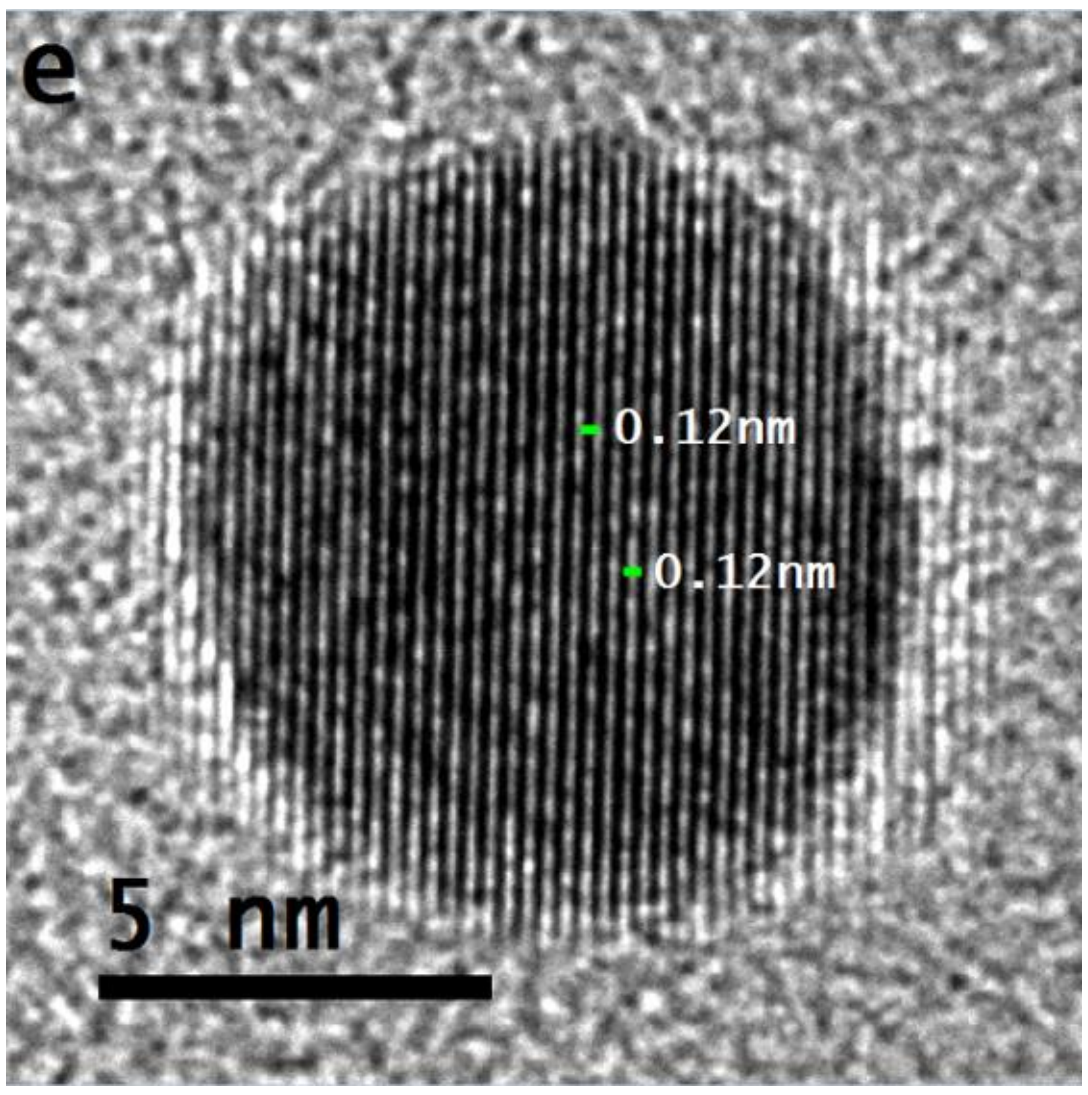

**Figure 2:** (a-e) HR-TOM images of tiny-sized particles; distance between the bottom of the copper capillary and solution surface ~ 0.5 mm, precursor concentration ~ 0.30 mM, and process duration ~ 30 sec

By increasing the distance between the bottom of the copper tube (the point of splitting argon atoms) and the solution surface, the electron streams of splitting argon gas atoms impinged to atoms underneath. In Figure 3 (a-c), atoms of tiny-sized particles dealt with different behavior of force and energy despite belonging to the same element. The distance between the bottom of the copper capillary and the solution surface was ~ 1.5 mm. In Figure 3 (a-c), some atoms of the tiny-sized particles also indicate the uniform stretching of energy knots. Atoms elongate under the adjacent orientations of electrons. In Figure 3 (a-c), tiny particles show different modes of amalgamation of atoms. Atoms mainly deal with deformation. So, the electronic structures of atoms did not deal with the exertions of even forces.

In Figure 3 (a), the exerted forces on the electrons do not convert the atoms into perfect structures of smooth elements. In Figure 3 (a), the atoms of the tiny-sized particle are less influenced by the traveling photons. The traveling photons along the air-solution interface do not keep a sufficient impact. In Figure 3 (a), atoms of the tiny-sized particle are less influenced by the traveling photons. In Figure 3 (b), some atoms of the tiny-sized particle elongate uniformly. In Figure 3 (c), a tiny-sized particle keeps the shape like an ellipse. Arrays of atoms do not fully convert into structures of smooth elements.

In Figure 3 (a-c), atoms of different tiny-sized particles deal with the exertions of forces in the solution but at a less pronounced level. Here, arrays of atoms do not convert into structures of smooth elements.

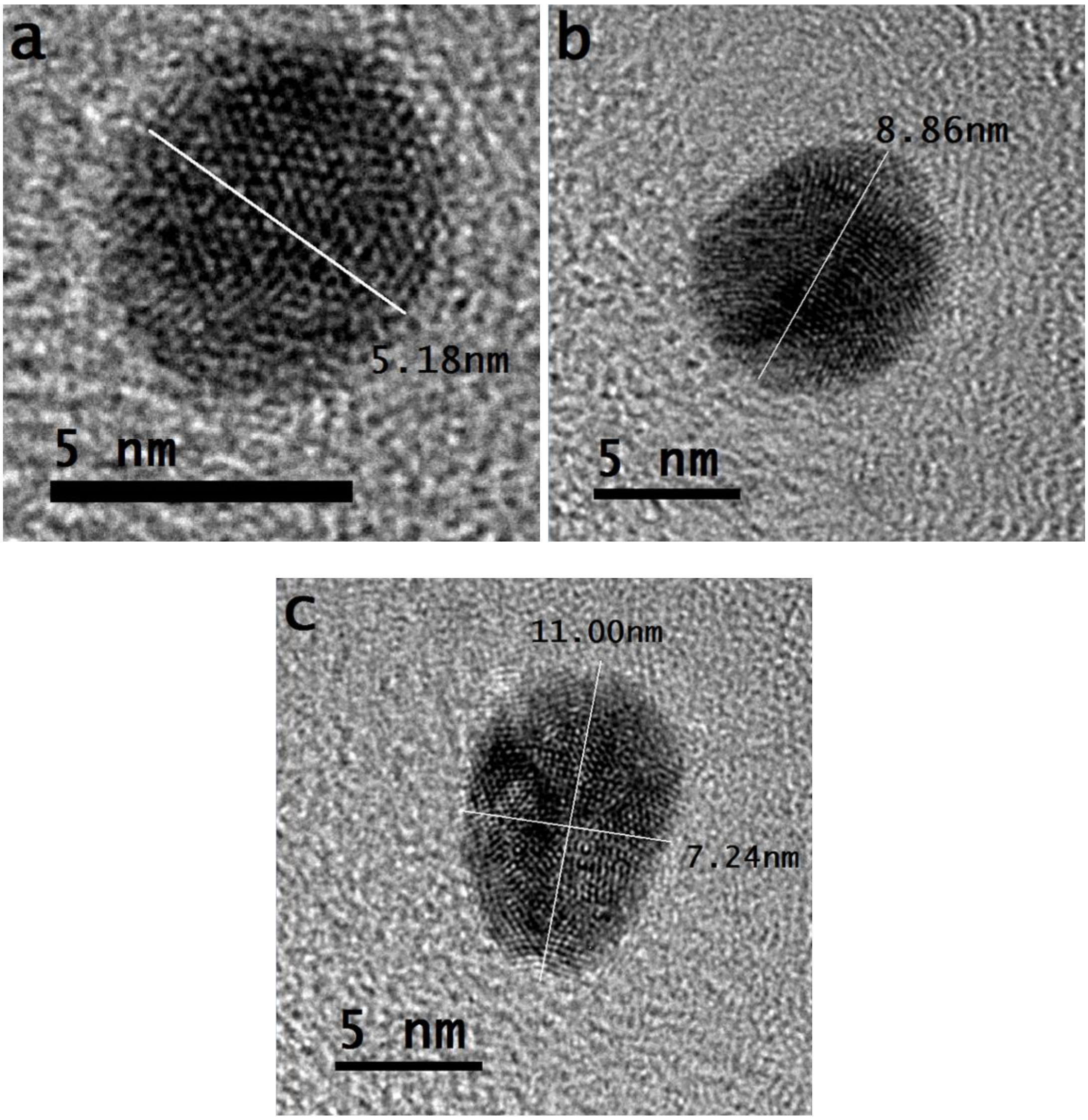

**Figure 3:** (a-c) HR-TOM images of tiny particles; distance between the bottom of the copper capillary and solution surface ~ 1.5 mm, precursor concentration ~ 0.30 mM, and process duration ~ 15 min

Increasing the precursor concentration from 0.30 mM to 0.40 mM increases the average size of the tiny particle. Different HR-TOM images in Figure 4 (a)-(c) reveal it. Figure 4 (a) shows all atoms of the tiny-sized particle undertook uniform stretching of the energy knot. Tiny clusters of atoms can also amalgamate to develop a tiny-sized particle.

Figure 4 (b) shows two portions of the tiny particle dealing force at the electron level. Arrows in Figure 4 (b) indicate those two portions. Arrays of atoms reveal uniform signs of stresses for half a tiny particle.

In Figure 4 (b), a half-portion of the tiny particle shows wrinkles, whereas the remaining half-portion does not. In Figure 4 (c), the tiny-sized particle shows the elongation of atoms from some portions, so atoms validate phase transitions. Atoms also validate different configurations of electronic structures.

Figure 4 (c) also indicates the original state of some atoms as they keep their spherical shapes. Most probably, those atoms executed confined interstate electron dynamics to bind. When the tiny particles develop in the solution, atoms do not deal with significant elongation. The exerted force on the electrons is in a less pronounced manner.

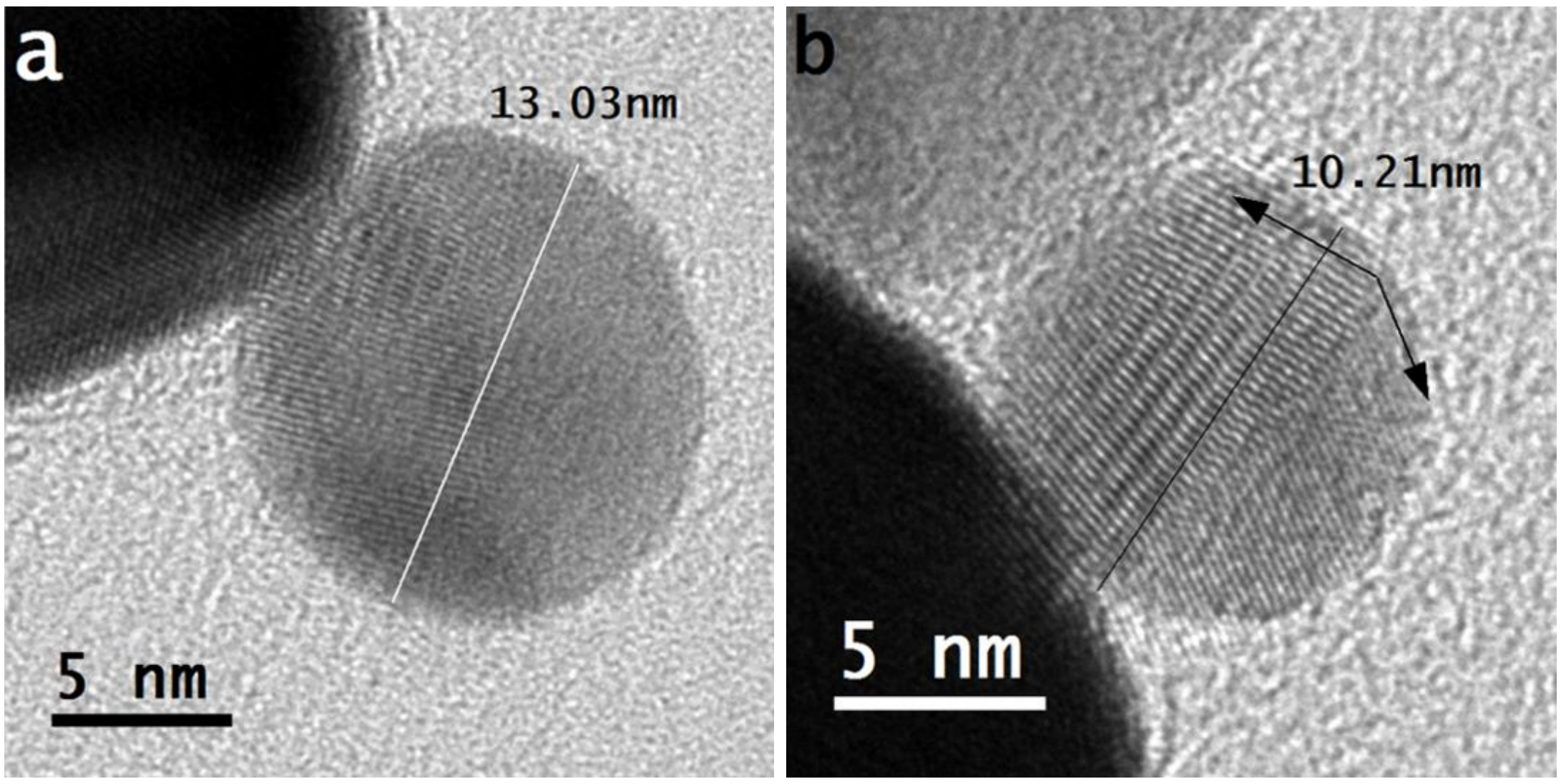

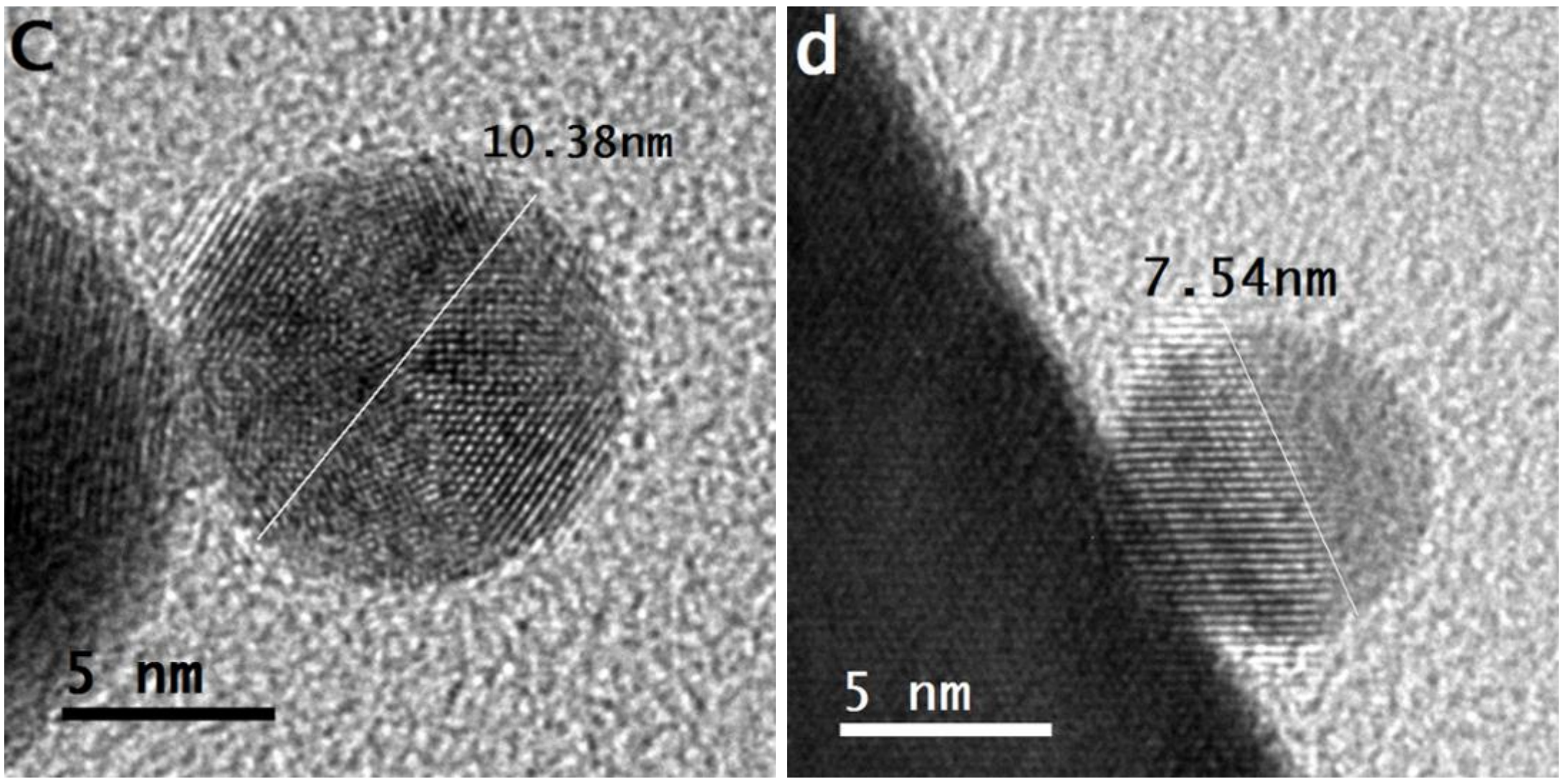

**Figure 4:** (a-d) HR-TOM images of the tiny particles; distance between the bottom of the copper capillary and solution surface ~ 1.0 mm, precursor concentration ~ 0.40 mM, and process duration ~ 20 min.

Atoms show different textures. More specifically, it is the case of the tiny-sized particle shown in Figure 4 (c). Figure 4 (d) also shows the atoms of the half-tiny particle elongated to develop structures of smooth elements. The smaller size of the tiny particle in Figure 4 (d) is due to the amalgamation of atoms at a later stage of the process.

Atoms deal with different configurations of electrons and phase transitions. It means a tiny particle deals with the transition state along the reaction path. So, the atoms also deal with the transition state along the reaction path. Here, a chemical aspect is more pronounced.

Hence, the specific phase transition of an atom is due to different forces and energy of electrons. Elongating atoms of arrays at an electronically decreasing level solution surface further shaped by the traveling photons [28]. However, structural evolution in the solid elements follows a different mechanism of binding atoms as the conservative forces involved to execute interstate electron dynamics [29]. A single atom of the mono-layer tiny particle elongates further when the impinging electron streams are at a fixed angle [30].

Atoms of mono-layer tiny particle deals with impinging electron streams from the external source. Gaseous and solid atoms undertake different transition states upon supplying energy [31].

Semisolid atoms execute interstate electron dynamics for generating photon energy [32]. In the pulse-based electron-photon/solution interface process, gold atoms dissociated from the precursor to develop different tiny-sized particles. An input source of the force and energy dissociates the precursor [33]. In addition to gold atoms, carbon atoms also amalgamate under attained dynamics [34-37].

Our results suggest that light and tiny particle interaction does not produce the collective oscillation as studied under the phenomenon of surface plasmons. According to this phenomenon, light (photons) travels along the air-solution interface, but traveling photons neither trap nor couple with the matter, i.e., atoms of tiny particles. These discussions support the study given elsewhere [28].

Again, gold atoms bind under a synthetic protocol to develop a tiny-shaped particle [28]. Gold atoms can also bind under the execution of electron dynamics [29]. Here, a structure should evolve instead of developing.

The van der Waals forces discussed elsewhere [38, 39] help to accomplish many studies. However, their roles now need to be re-visited. They do not justify the binding mechanisms at full scale. Nevertheless, the authors greatly commend the investigations. The basic idea of binding atoms in different states of carbon atoms is also discussed [40]. It is different from the binding of gold atoms.

A study published elsewhere [41] discussed the binding in gold atoms and their dynamics. Some more studies target the hexagonal close-packed structure [12] and geometric structure [12, 42]. In colloidal matter, ongoing research efforts not only consider the structure to explain entropy and geometry but also dynamics [43]. However, to master the new insights and advances in colloidal science, the studies given elsewhere [28, 41] are vital.

In developing the tiny-sized particles, atoms can deal with different dynamics, different configurations of electronic structure, and different phase transitions. Due to miscellaneous interactions, an early-stage developed tiny particle can distort. Atoms of arrays elongate under various interactions. So, their structures of smooth elements can deteriorate.

In tiny particles of non-geometric shapes, atoms can deal with different configurations of electronic structure and phase transitions. Electrons unevenly deal with forces. The electronic structures of the gold atoms alter in their tiny particles. It is mainly because of the process of synergy, impinged electron streams and interacted photons.

Tiny-sized particles develop under different modes of electronic structure, phase transition, and localized dynamics of atoms, so they disregard the phenomenon of surface plasmons. The developmental process of the tiny-sized particle partially obeys the van der Waals' forces to bind atoms since the element of energy is also required. The study of nanoscale components opens new horizons to science.

## 4.0. Conclusion

Tiny-sized particles develop in different shapes depending on the attained dynamics of atoms. Results show atoms amalgamate under different dynamics for each tiny particle. A tiny-sized particle does not keep any specific geometry when its atoms amalgamate with non-uniform dynamics.

Different factors contribute to determining the size of a tiny particle. Atoms deal with different dynamics to develop a tiny-sized particle. So, many studies are needed to conclude any result.

Atoms of tiny-sized particles undertake different transition states. Atomic electrons do not keep equal potential energy and the same orientation. When tiny-sized particles develop by the groups of atoms, each group of atoms amalgamates under significant dynamics.

Developing tiny particles in the solution does not keep specific shapes. Supplied energy is the source of binding atoms. Some atoms can bind through the process of synergy. Some atoms can also bind under the execution of interstate electron dynamics. The development of tiny particles here validates the role of localized dynamics.

A tiny particle can decrease the width of the structure of a smooth element. Localized dynamics of atoms deal with both medium-based dynamics and externally influencing dynamics.

A tiny particle of gold can not evolve structure at the solution surface. The main contribution is related to the surface force. However, some atoms can bind in the solution. In structural evolution, all atoms of a tiny particle should execute electron dynamics, as discussed elsewhere [29]. Overall, tiny-sized particles developed at different processing parameters show major level modifications of their structures.

**Acknowledgement:**

This study was funded by the National Science Council, now the Ministry of Science and Technology, Taiwan (grant number NSC-102-2811-M-032-008; 2013-2014). The authors wish to thank Mr. Chien-Jui Yeh for assisting in the microscopic operation.

## Authors’ biographies:

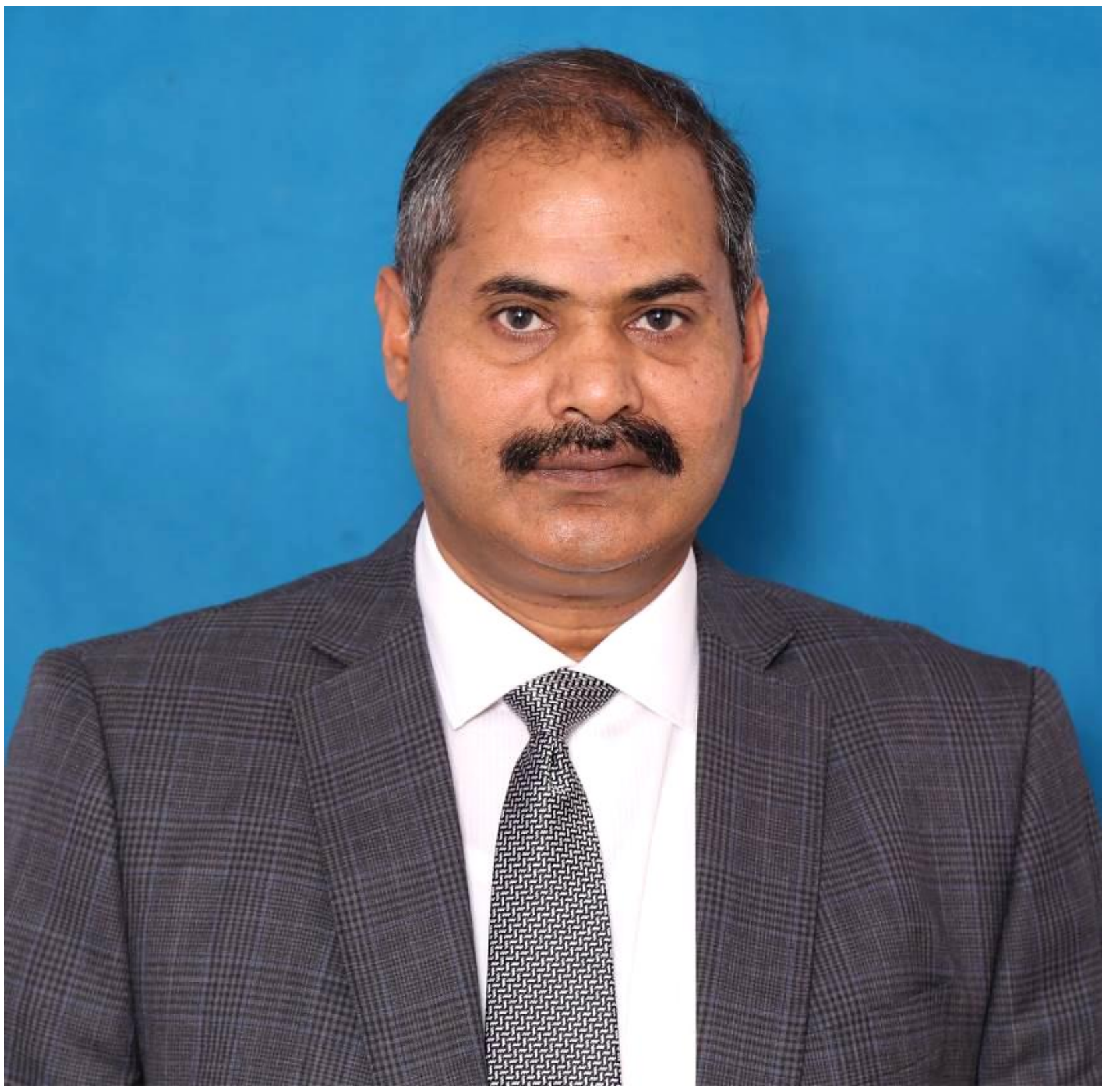

In 1996, **Mubarak Ali** earned a B.Sc. degree in Physics and Mathematics. The University of the Punjab awarded him the degree. The M.Sc. degree in Materials Science got in 1998. Bahauddin Zakariya University Multan awarded him the degree of master with distinction. He completed his M.Sc. thesis at Quaid-i-Azam University Islamabad. He gained a PhD in Mechanical Engineering from the Universiti Teknologi Malaysia under the award of the Malaysian Technical Cooperation Programme (MTCP;2004-07) and a postdoc in advanced surface technologies at Istanbul Technical University under the foreign fellowship of The Scientific and Technological Research Council of Turkey (TÜBİTAK, 2010). Dr Mubarak completed another postdoc in nanotechnology at the Tamkang University Taipei, 2013-2014, sponsored by the National Science Council, now the Ministry of Science and Technology, Taiwan. He remained working as an Assistant Professor on the tenure track at COMSATS University Islamabad from May 2008 to June 2018, previously known as COMSATS Institute of Information Technology. His new position is in process. Before that, he remained a working assistant and deputy director at M/o Science & Technology, Pakistan Council of Renewable Energy Technologies, Islamabad, from January 2000 to May 2008. The Institute for Materials Research at Tohoku University Japan invited Dr. Mubarak to deliver a scientific talk. His scientific research remained a part of many conferences organized by renowned universities in many countries. His core area of research includes materials science, physics & nanotechnology. He also won a merit scholarship for PhD study from the Higher Education Commission, Government of Pakistan. However, he did not obtain this opportunity. He earned a diploma (in English) and a certificate (in the Japanese language) in 2000 and 2001, respectively, part-time from the National University of Modern Languages, Islamabad. He is the author of several articles. Please refer to the link
https://www.researchgate.net/profile/Mubarak_Ali5
and the link https://scholar.google.com.pk/citations?hl=en&user=UYjvhDwAAAAJ

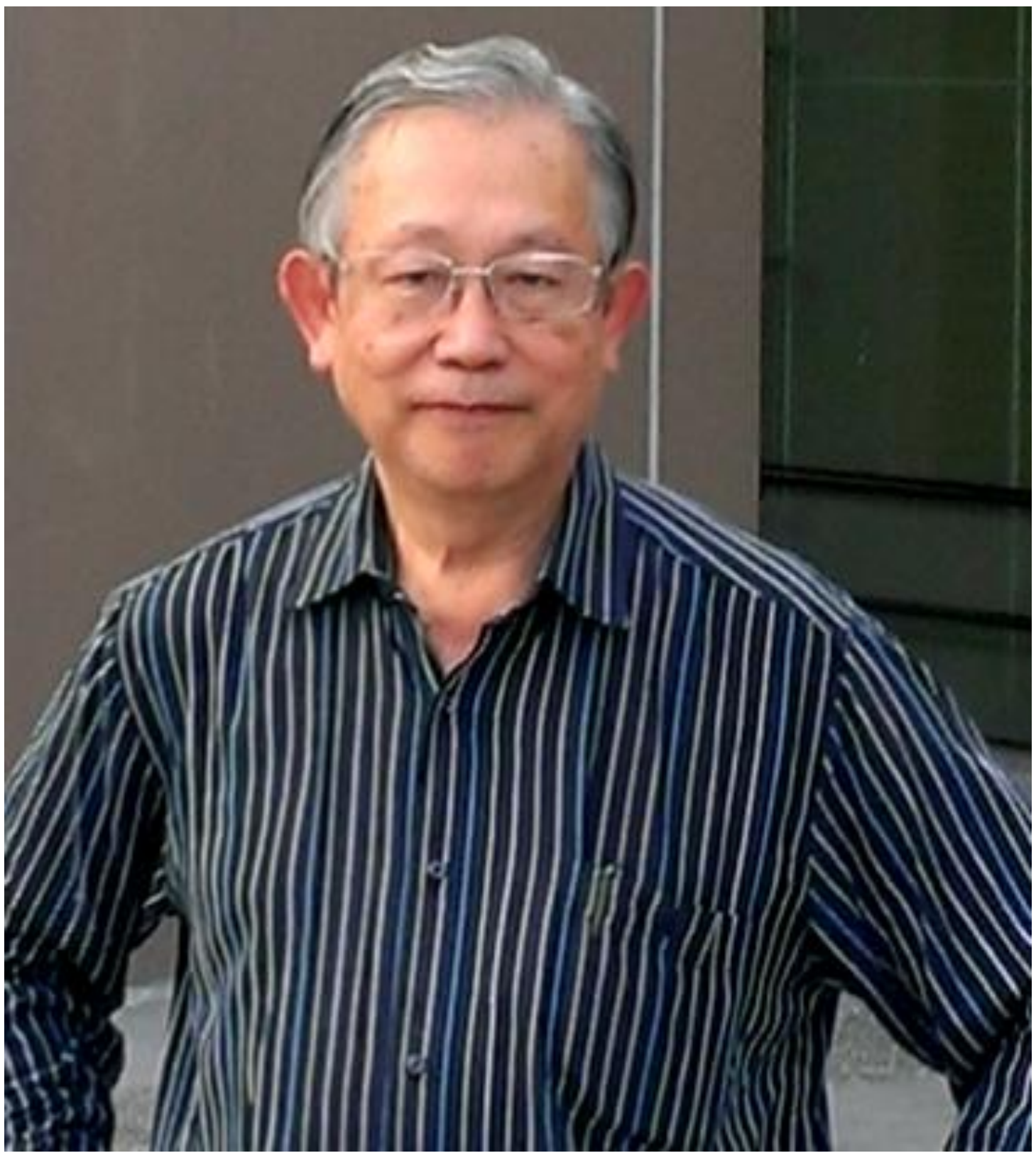

His Excellency, **I-Nan Lin**, worked as a **Senior Professor** for several years at Tamkang University, Taiwan, and is now **Professor Emeritus**. His Excellency received a bachelor's degree in physics from National Taiwan Normal University, Taiwan, an M.S. from National Tsing-Hua University, Taiwan, and a Ph.D. degree in Materials Science from U. C. Berkeley in 1979, U.S.A. His Excellency worked as a senior researcher in the Materials Science Centre at Tsing-Hua University for several years and as a faculty member and senior professor in the Department of Physics at Tamkang University. Professor Lin published several hundred referred journal publications and held a distinctive position in the university. Professor I-Nan Lin supervised several PhD and Postdoc candidates around the world. Professor I-Nan Lin continued researching in the areas of developing high-conductivity diamond films, composite materials, and deep-rooted studies of the transmission microscopy of materials.